\documentclass[twocolumn,showpacs,twoside,10pt,superscriptaddress,prl]{revtex4}
\usepackage{graphicx}
\usepackage{epsfig}
\usepackage{epsf}
\usepackage{amssymb}
\usepackage{amsmath}

\begin{document}

\title{Is the quantum adiabatic theorem consistent?}
\author{Jiangfeng Du}
\email{djf@ustc.edu.cn}
\affiliation{Hefei National Laboratory for Physical Sciences at Microscale, Hefei, Anhui
230026, People's Republic of China}
\affiliation{Department of Modern Physics, University of Science and Technology of China}
\author{Lingzhi Hu}
\affiliation{Hefei National Laboratory for Physical Sciences at Microscale, Hefei, Anhui
230026, People's Republic of China}
\author{Ya Wang}
\affiliation{Hefei National Laboratory for Physical Sciences at Microscale, Hefei, Anhui
230026, People's Republic of China}
\affiliation{Department of Modern Physics, University of Science and Technology of China}
\author{Jianda Wu}
\affiliation{Hefei National Laboratory for Physical Sciences at Microscale, Hefei, Anhui
230026, People's Republic of China}
\affiliation{Department of Modern Physics, University of Science and Technology of China}
\author{Meisheng Zhao}
\affiliation{Hefei National Laboratory for Physical Sciences at Microscale, Hefei, Anhui
230026, People's Republic of China}
\affiliation{Department of Modern Physics, University of Science and Technology of China}
\author{Dieter Suter}
\affiliation{Fachbereich Physik, Universit\"{a}t Dortmund, 44221 Dortmund, Germany}
\date{\today}

\begin{abstract}
The quantum adiabatic theorem states that if a quantum system starts in an
eigenstate of the Hamiltonian, and this Hamiltonian varies sufficiently
slowly, the system stays in this eigenstate. We investigate experimentally
the conditions that must be fulfilled for this theorem to hold. We show that
the traditional adiabatic condition as well as some conditions that were
recently suggested are either not sufficient or not necessary. Experimental
evidence is presented by a simple experiment using nuclear spins.
\end{abstract}

\pacs{03.65.Ca, 03.65.Ta, 76.60.-k} \maketitle

In classical physics, adiabatic processes do not involve a transfer of heat
between system and environment. In quantum mechanics, the adiabatic theorem
states that a system that is initially in an eigenstate of the Hamiltonian
will remain in this eigenstate if the changes of this Hamiltonian are
sufficiently slow \cite{Ehrenfest,Born,Kato,Messiah}. While this \emph{%
adiabatic theorem} (QAT) is a well established fact, it appears to be
difficult to formulate a consistent \emph{adiabatic condition} (QAC), which
unambiguously states when the theorem applies and is both necessary and
sufficient.

The QAT is critical for the booming of many domains in quantum mechanics. It
provides the foundation and insightful interpretation of Landau-Zener
transition \cite{Landau}, the Gell-Mann-Low theorem \cite{Gell} and Berry's
phase \cite{Berry}. Quantum adiabatic processes are also used for some
quantum algorithms \cite{Farhi,Steffen}, which can efficiently solve
NP-complete problems. These algorithms are based on the validity of the QAT
and a sufficient QAC \cite{Roland}.

Recently, however, doubts were cast over the consistency of the QAT and the
sufficiency of the QAC. Marzlin and Sanders first suggested a possible
inconsistency of the QAT \cite{Marzlin1}. Although there are some
questionable points in their deduction \cite{Marzlin2,Marzlin3,Marzlin4},
their main point resulted in an extended discussion \cite{Wu1}. Then Tong
\emph{et al} gave a specific counterexample to show that the traditional QAC
is not sufficient for the adiabatic approximation to hold \cite{Tong1}.
These discussions about QAT and QAC resulted in further investigations such
as modification of traditional QAC \cite{Tong2}, reexamination of the
quantum adiabatic algorithm \cite{Wei}, and study of QAC in different
quantum systems \cite{Sarandy}. While there is fast progress in the
theoretical discussion about QAC and QAT, an unambiguous experimental
investigation is certainly important here. However, such experiements still
remained a real challenge due to the following reasons: $\left( 1\right) $
the conflict between the sufficiently long time during the adiabatic
evolution of the time-dependent Hamiltonian in QAT and the severely short
coherent time of the real physical system due to the decoherence. $\left(
2\right) $ the suitable technique with good quantum controlling during the
quantum adiabatic process.

Considering that the coherent time of nuclei spin inside the atom is
relatively longer compared to that of other physical systems and
nuclear magnetic resonance (NMR) is well developed over the past
decades, here we first present a simple and clear-cut experimental
investigation of this issue, using a spin-1/2 particle in a rotating
magnetic field. We show that, depending on the parameters chosen,
the traditional QAC is neither sufficient nor necessary. Then we
theoretically compare various newly proposed QAC's with the
traditional one and compare their applicability to our specific
system. We also provide further experimental proof to support our
theoretical comparison and discuss the character of the different
adiabatic conditions.

The quantum adiabatic theorem states that if the energy levels of a time
dependent Hamiltonian $H(t)$ are never degenerate and the Hamiltonian varies
sufficiently slowly with time, the initial eigenstate of this Hamiltonian
will stay close to the instantaneous eigenstate at a later time \cite%
{Messiah}.

The widely used qualitative condition that assures the QAT valid is the QAC:
\begin{equation}
|\frac{\langle E_{m}(t)|\dot{E}_{n}(t)\rangle}{E_{m}(t) - E_{n}(t)}|\ll1,
\quad m\neq n,\quad t\in[0,T],
\end{equation}
where $E_{m}(t)$ and $|E_{m}(t)\rangle$ are the instantaneous eigenvalues
and eigenstates of $H(t)$, and $T$ is the total evolution time. We define
the fidelity as the absolute value of the overlap of the actual state and
the instantaneous eigenstate: $F(t)=|\langle\Psi(t)|\phi(t)\rangle|$, where $%
|\Psi(t)\rangle$ is the instantaneous eigenstate of the Hamiltonian and $%
|\phi(t)\rangle$ is the state that has evolved under the Hamiltonian $H(t)$
from $|\Psi(0)\rangle$. With this definition, the adiabatic theorem can be
formulated such that the fidelity F(t) will stay close to 1 if the variation
of Hamiltonian meets condition (1).

As a specific Hamiltonian, we choose
\begin{equation}
H(t)=\omega_{0}\frac{\sigma_{z}}{2}+\omega_{1}(\frac{\sigma_{x}}{2}%
\cos\omega^{\prime}t+\frac{\sigma_{y}}{2}\sin\omega^{\prime}t)
\end{equation}
where $\omega_{0}$ is the Larmor frequency, $\omega_{1}$ is the strength of
the coupling to a radio frequency(rf) magnetic field, and $\omega^{\prime}$
is the rotation frequency of the rf magnetic field. We investigate the
validity of the adiabatic theorem as a function of the strength and
frequency of the rf field.

Experiments were performed on $^{13}$C-labeled CHCl$_{3}$ at room
temperature using a Bruker AV-400 spectrometer. The experiments were
performed on the $^{13}$C nuclear spin, while the $^{1}$H nuclear spin was
decoupled during the whole experiment.

For the sake of convenience, we define two parameters: $R=\omega_{1}/%
\omega_{0}$ and $K=\omega^{\prime}/\omega_{0}$. In our first experiment, we
choose the power of the rf field $\omega_{1}=100Hz$ and $\omega_{0}=1700Hz$,
corresponding to $R=0.06\ll1$.

The initial Hamiltonian $H(0)$ has an eigentstate $|\Psi (0)\rangle =\cos
\frac{\theta }{2}|0\rangle +\sin \frac{\theta }{2}|1\rangle $, in which $%
\theta =\arctan R=\arctan 0.06$. We can prepare this initial state by
applying an rf pulse along the y-axis, with a rotation angle $\arctan 0.06$,
to the thermal equilibrium state $|0\rangle $.

To realize an evolution determined by the Hamiltonian (2), we use the
discrete approach proposed by Steffen \cite{Steffen}. The rotation of the rf
field, at frequency $\omega ^{\prime }$, was performed by applying a
sequence of small flip-angle pulses, whose phase was initially set to zero
and shifted by $\frac{\pi }{36}$ for every pulse.

We consider two specific cases: $K=1$ and $K=10$ in the following. It is
easy to prove that in the first case the evolution under the Hamiltonian (2)
satisfies the adiabatic condition (1), while in the second case the
evolution of the Hamiltonian violates the adiabatic condition seriously.

We first consider the case $K=1$, which means $\omega ^{\prime }=\omega
_{0}=1700Hz$. Thus the width of each flip-angle pulse can be calculated as $%
\Delta t=\frac{\left( \pi /36\right) }{2\pi \omega ^{\prime }}=8.2\mu s$. We
can compute the time that the rf field rotates one circle as $\tau =\frac{%
2\pi }{\pi /36}\cdot \Delta t=590.4\mu s$. We measure the state of
the spin after it evolves $n$ circles, in other word, at the time
$t=n\tau $, in which $n$ changes from form $0$ to $15$. We can
calculate the fidelities at these time points and the experimental
results are summarized as black circles in Fig. 1.

By simply changing the pulse width $\Delta t$ in the above experiment, we
can realize the case $K=10$. Here, $\omega ^{\prime }=10\omega _{0}=17000Hz$
and $\Delta t=0.82\mu s$. We repeat the same process in the last experiment
and the result is presented as red squares in Fig. 1.

\begin{figure}[tbp]
\epsfig{file=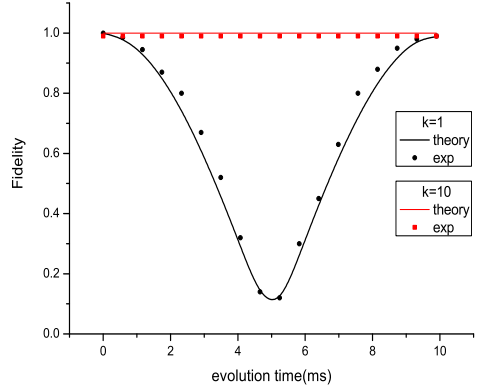,width=7cm,height=6cm} \caption{(color online)
Measured fidelity when $K=1$ compared to the fidelity when $K=10$.
The black curve and red curve are the theoretical results of $K=1$
and $K=10$ respectively. The black circles and red squares are the
experimental results of $K=1$ and $K=10$ respectively.}
\end{figure}

An interesting and exciting phenomenon is that when $K=1$ and the
traditional adiabatic condition is satisfied, the state evolves far away
from instantaneous eigenstate and the fidelity falls below $0.1$ at $t=5ms$.
Therefore, we can conclude that the adiabatic condition is not sufficient.
On the other hand, when $K=10$, even though the traditional adiabatic
condition is violated, the state is always next to the instantaneous
eigenstate and the fidelity remains close to $1$. So the adiabatic condition
(1) is not necessary. Synthesizing these two cases, it is evident that the
traditional QAC is indeed problematic.

\begin{figure*}[tbp]
\epsfig{file=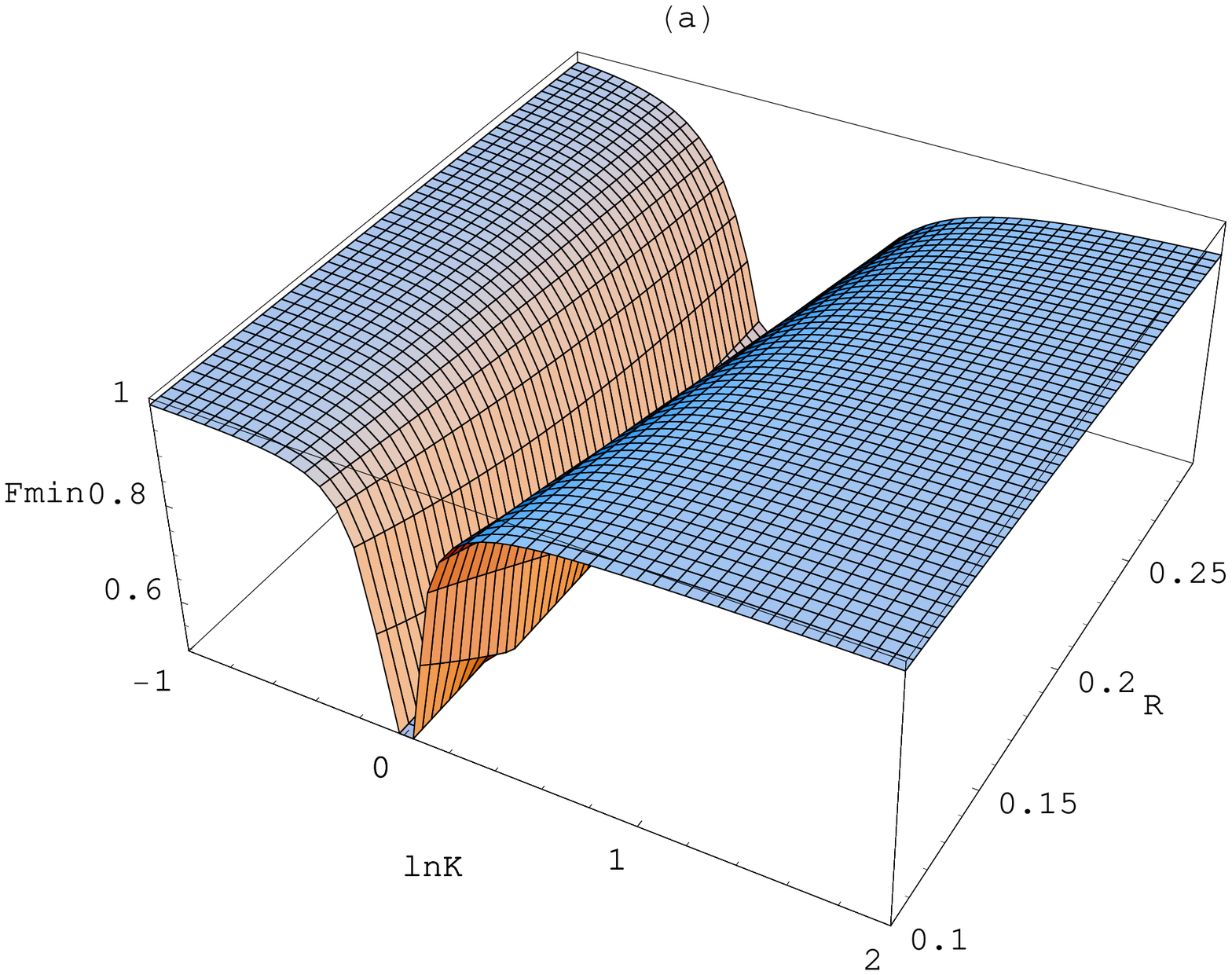,width=6cm,height=5.5cm}
\epsfig{file=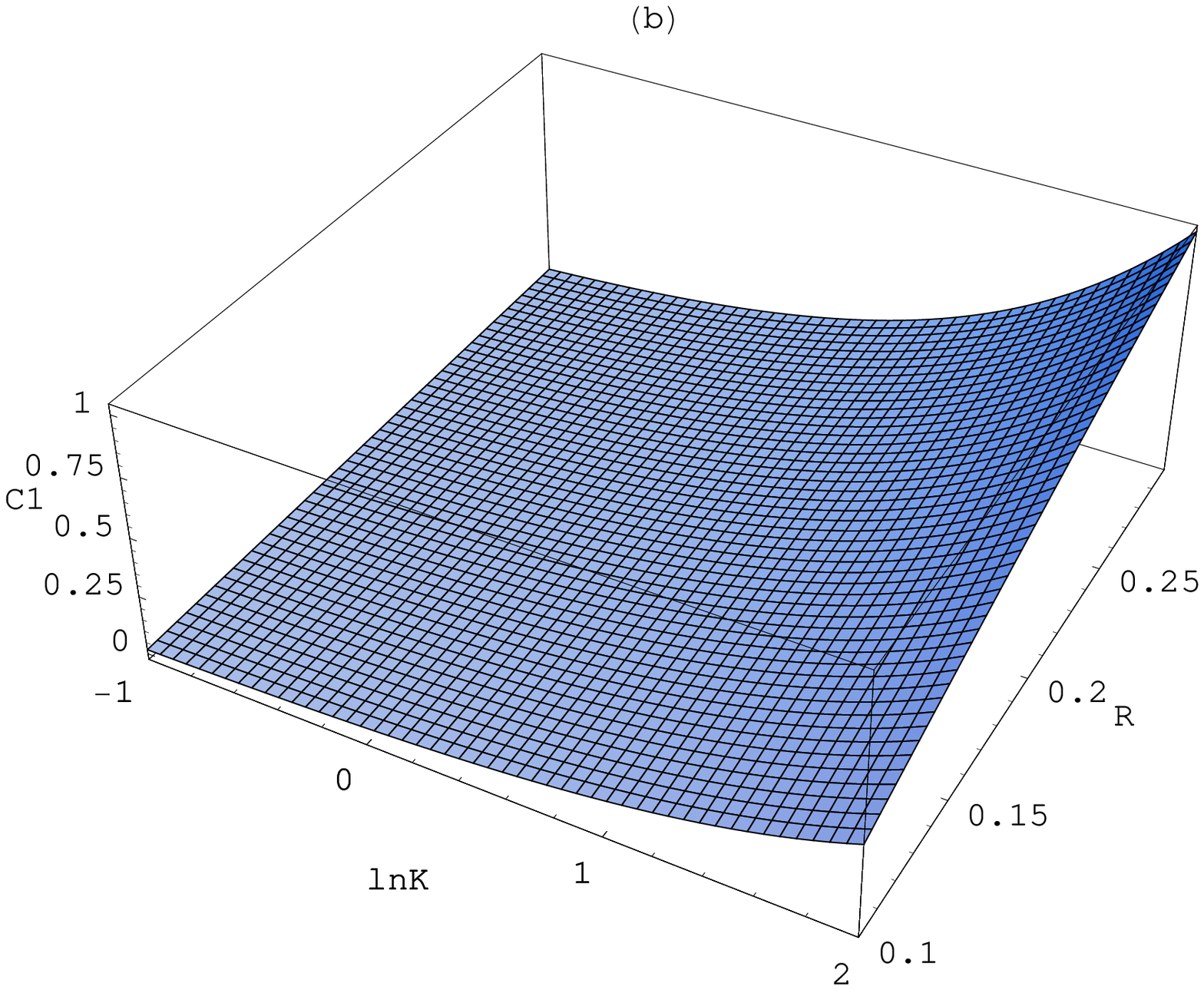,width=6cm,height=5.5cm}
\epsfig{file=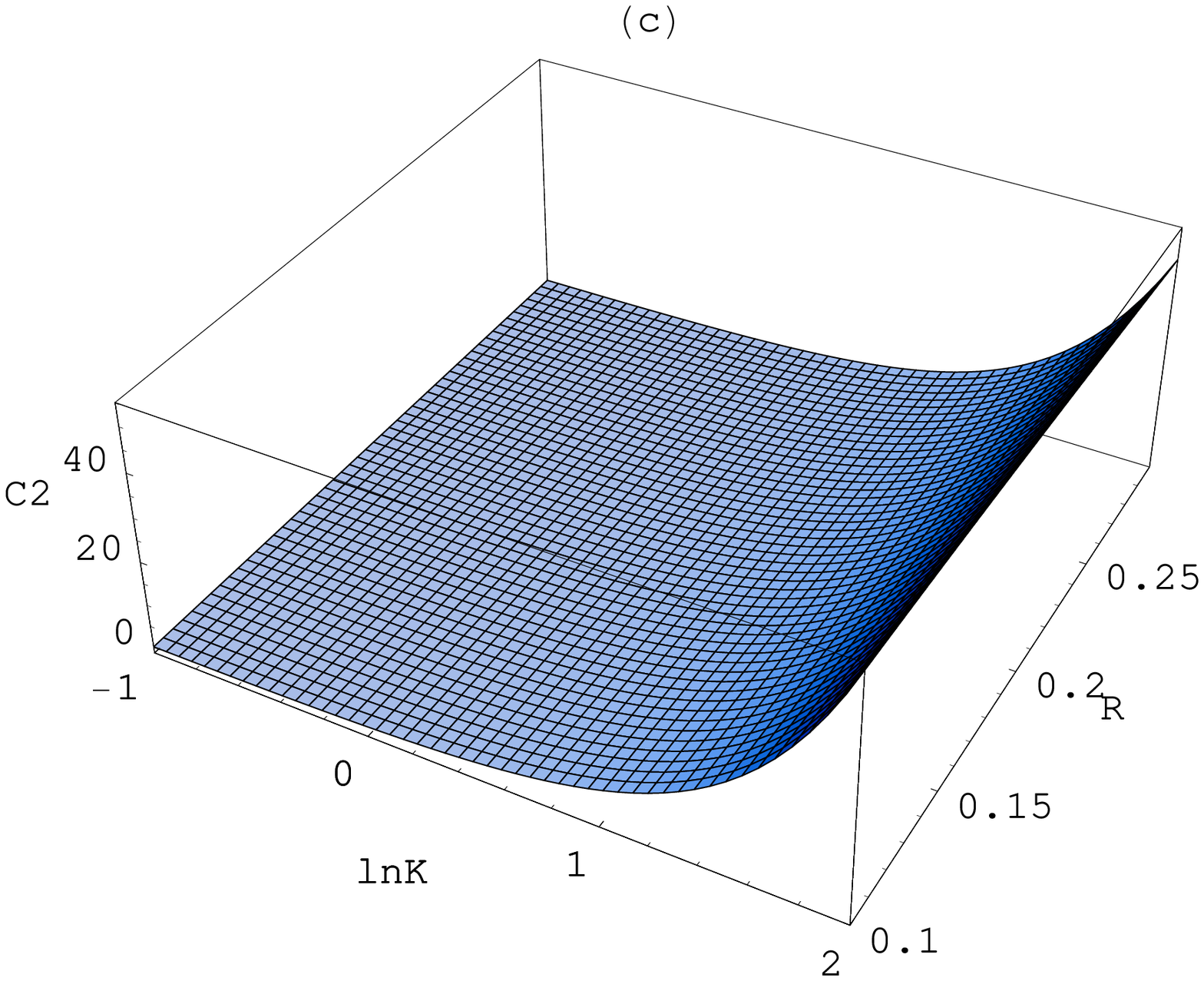,width=6cm,height=5.5cm}
\epsfig{file=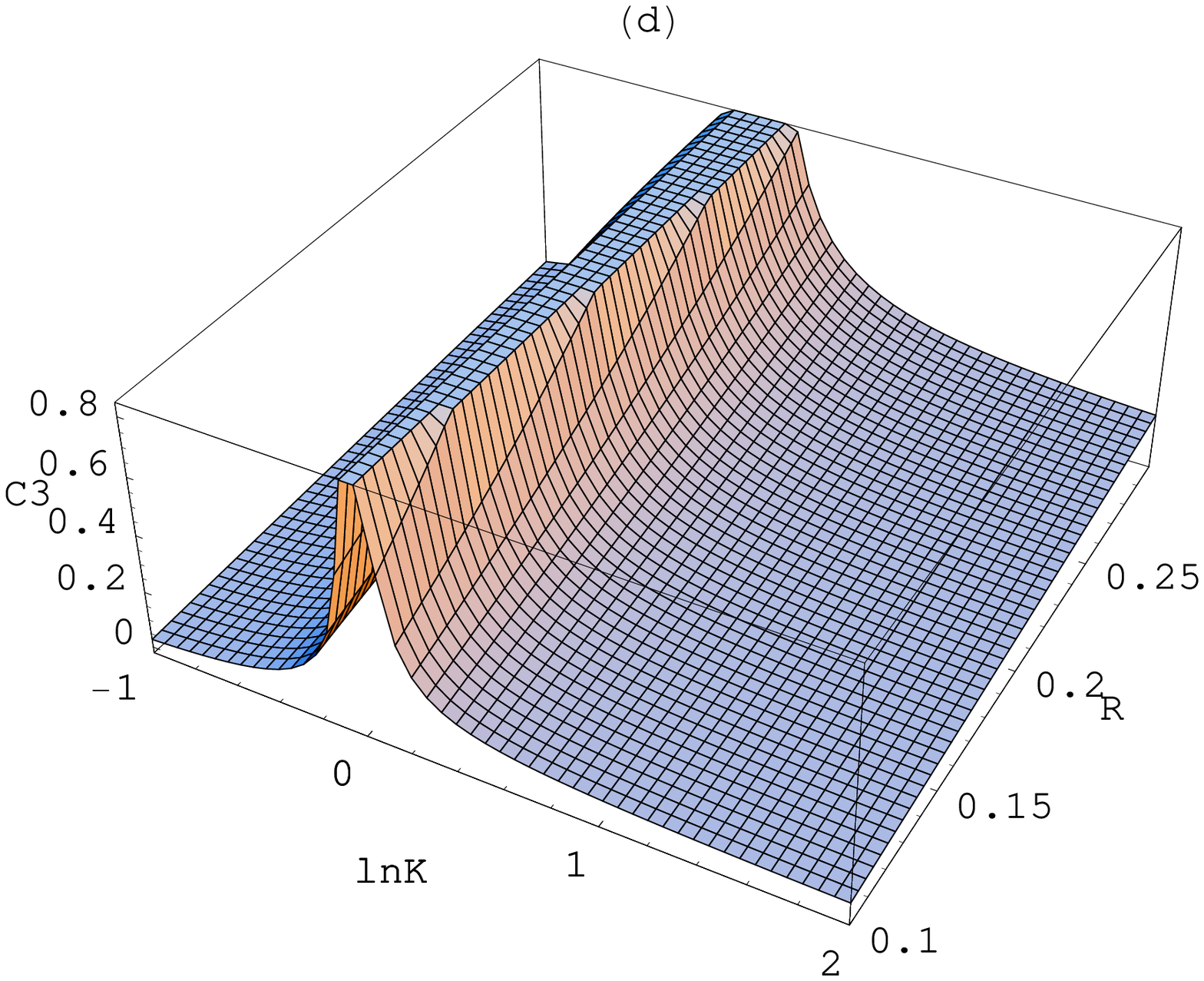,width=6cm,height=5.5cm} \caption{(color
online) (a) The minimum fidelity in the process of evolution as a
function of $lnK$ and $R$. (b)-(d) show the traditional QAC, Tong's
and Wu's expression as a function of the same parameters.}
\end{figure*}

Next, we examine the validity of other, more recently proposed adiabatic
conditions, again using the Hamiltonian (2). We compare the traditional QAC,
Tong's QAC and Wu's QAC \cite{Tong2}. Analytically solving the Schr\"{o}%
dinger equation, we can calculate $F_{min}$, the minimal fidelity of F(t) in
the process of evolution:
\begin{equation}
F_{min}=\frac{(1-K)\cos \theta +R\sin \theta }{\sqrt{(1-K)^{2}+R^{2}}}
\end{equation}%
where $\theta $ was defined earlier. For the sake of convenience, we define $%
C_{1}$ as the expression in traditional QAC: $C_{1}=|\frac{\langle E_{+}(t)|%
\dot{E}_{-}(t)\rangle }{E_{+}-E_{-}}|$.

Using fundamental inequalities, Tong \emph{et al} states that the adiabatic
approximation will be reasonable if the Hamiltonian satisfies the following
conditions
\begin{eqnarray}
&(A)& \quad |\frac{\langle E_{m}(t)|\dot{E}_{n}(t)\rangle}{E_{m}-E_{n}}%
|\ll1,\quad t\in[0,T], \\
&(B)&\quad \int_{0}^{T}|(\frac{\langle E_{m}(t)|\dot{E}_{n}(t)\rangle}{%
E_{m}-E_{n}})^{\prime}|dt\ll1, \\
&(C)& \quad \int_{0}^{T}|\frac{\langle E_{m}(t)|\dot{E}_{n}(t)\rangle\langle
E_{n}(t)|\dot{E}_{l}(t)\rangle}{E_{m}-E_{n}}|dt\ll1,
\end{eqnarray}
in which $\langle m(t)|\dot m(t) \rangle=0$, $m\neq n$, $n\neq l$, and $T$
is the total evolution time. For our system, condition (5) is the strongest
of these. Therefore we define $C_{2}=\int_{0}^{T}|(\frac{\langle E_{+}(t)|%
\dot{E}_{-}(t)\rangle}{E_{+}-E_{-}})^{\prime}|dt$, in which $T$ equals $%
1/\omega^{\prime}$, a typical evolution time.

On the basis of invariant perturbation theory, Wu \emph{et al} deduced a
modified adiabatic condition
\begin{equation}
|\frac{\langle E_{m}(t)|\dot{E}_{n}(t)\rangle}{E_{m}-E_{n}+\Delta_{nm}}%
|\ll1,\quad t\in[0,T].
\end{equation}
In the condition (7)
\begin{eqnarray}
\Delta_{nm}=i\langle E_{n}(t)|\dot{E}_{n}(t)\rangle-i\langle E_{m}(t)|\dot{E}%
_{m}(t)\rangle  \notag \\
+i\frac{d}{dt}\arg\langle E_{m}(t)|\dot{E}_{n}(t)\rangle
\end{eqnarray}
is $U(1)$-invariant under the time-dependent transformation, and it is just
the difference of Berry phase between different evolution orbits if it is
integrated along a cycle. For the Hamiltonian (2), we can rewrite the
condition (7) as $C_{3}=|\frac{\langle E_{+}(t)|\dot{E}_{-}(t)\rangle}{%
E_{+}-E_{-}+i\langle E_{-}(t)|\dot{E}_{-}(t)\rangle-i\langle E_{+}(t)|\dot{E}%
_{+}(t)\rangle}|$.

Figure 2 summarizes $F_{min}$, $C_{1}$, $C_{2}$ and $C_{3}$ as functions of $%
K$ and $R$. In each case, the adiabatic condition $C_{i}\ll1$ should imply $%
F_{min}\approx1$ ($i=1,2,3$). For the traditional QAC (Fig. 2b), we
note that at $K \approx 1$, the QAC is fulfilled, but $F_{min}$
falls well below $1$. Conversely, when $K \gg 1$, $F_{min}$ remains
close to 1. Apparently, the traditional QAC is neither necessary nor
sufficient.

Similarly, adiabatic condition $C_{2}\ll1$ is fulfilled only for $K\ll1$.
Because $F_{min}\approx1$ if $K\ll1$ or $K \gg 1$ (When $R\ll1$), we can
conclude that this adiabatic condition is sufficient but not necessary. The
adiabatic condition most suitable to our Hamiltonian is $C_{3}\ll1$. $C_{3}$
is small where $F_{min}$ is close to 1 and vice versa. For our system, Wu's
adiabatic condition is therefore sufficient as well as necessary.

We have experimentally verified the calculations represented in Fig. 2. In
the experiment, we measured the minimum fidelity as a function of $K$ at
fixed $R$ and as a function of $R$ at fixed $K$. In this experiment, the
average rf field strength was $\omega_{1}=100$ Hz, and we used the same
discrete method for the implementation of the time-dependent Hamiltonian as
in the first experiment. We changed $K$ from $0.5$ to $1.5$ by varying the
width of the flip-angle pulses, and we varied $R$ from $0.05$ to $0.3$ by
varying the frequency offset $\omega_0$. The most important difference from
the first experiment is that here we did not measure the state after a
cyclic evolution, at $t=n\tau$, but at the time of the minimum fidelity, $%
t_{min}=\frac{1}{2\sqrt{(1-K)^{2}\omega _{0}^{2}+\omega _{1}^{2}}}$. As an
example, for the parameters $R=0.05$ and $K=0.75$, $t_{min}=980.6\mu s$. The
experimental points were not chosen equidistant as a function of $K$, but
denser around $K=1$, where the minimum fidelity changes rapidly. The
experimental results are represented in Fig. 3; obviously, the agreement
with the theoretical predictions of Fig. 2 (represented as the curves in
Fig. 3) is more than satisfactory.

\begin{figure}[tbp]
\epsfig{file=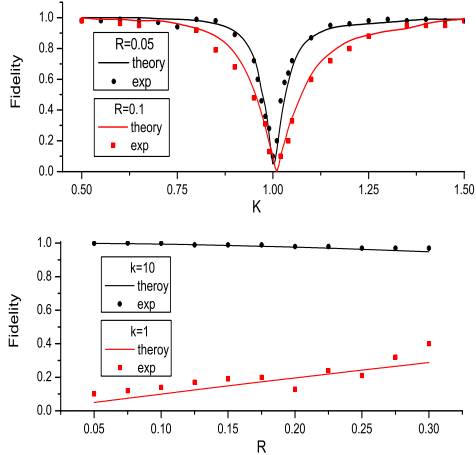,width=7cm,height=7cm} \caption{(color online)
(a) The fidelity during the evolution measured as a function of $K$.
The black circles and red squares are the experimental minimum
fidelities for $R=0.05$ and $R=0.1$ respectively. The black and red
curves are the theoretical results. (b) The minimum fidelity as a
function of $R$. The black circles and red squares represent the
experimental data for $K=10$ and $K=1$ respectively. The black and
red curves are the theoretical results. }
\end{figure}

Qualitatively, the observed behavior for the case of $K=1$ can be
understood as a resonant phenomenon. Although the perturbation (rf
in our experiment) is very small, it can seriously affect the
evolution if it contains a frequency component that matches a
transition frequency of the system. The traditional QAC and Tong's
condition do not account for resonant effects, but in Wu's adiabatic
condition takes includes it's effect.

In this experiment, the parameter $R$ reflects the angle between the
rotating magnetic field and the z axis. Fig. 3 demonstrates that if the
magnetic field is very close to the z axis, the resonance region is narrow.
If the angle increases ($R$ increases), the "resonance" becomes both wider
and deeper. If we consider the behavior as a function of $R$, we observe
different behaviors, depending on $K$.

Several adiabatic algorithms have been proposed for quantum computing; in
most cases, their validity was discussed in terms of the traditional QAC
\cite{Roland}. Since this condition is neither sufficient nor necessary, the
validity of these adiabatic algorithms should be re-evaluated. Conversely,
if they are designed to fulfill a QAC that is not necessary, these
algorithms may not realize their full potential. Although Wu's QAC seems to
be proper in our specific system, it is still not always a sufficient and
necessary condition \cite{Tong2}. Therefore, finding a QAC that is both
sufficient and necessary remains an open question. Finding such a condition
will be important for the development of quantum adiabatic algorithms.

In conclusion, we have demonstrated that the traditional adiabatic condition
is neither sufficient nor necessary. We found that the most important
deviation can be understood as a resonant effect. Rather than concluding
that the quantum adiabatic theorem is inconsistent \cite{Marzlin1}, we hold
that it is important to determine the correct condition for an adiabatic
evolution of a quantum system.

\begin{acknowledgments}
We thank Yongde Zhang for helpful discussion. Financial support comes from
National Nature Science Foundation of China, the CAS, Ministry of Education
of PRC, and the National Fundamental Research Program. It is also supported
by Marie Curie Action program of the European Union.
\end{acknowledgments}


\begin{thebibliography}{99}
\bibitem{Ehrenfest} P. Ehrenfest, \textit{Ann. Phys} \textbf{356}, 327
(1916).

\bibitem{Born} M. Born and V. Fock, \textit{Z. Phys.} \textbf{51}, 165
(1928).

\bibitem{Kato} T. Kato, \textit{J. Phys. Soc. Jpn.} \textbf{5}, 435 (1950).

\bibitem{Messiah} A. Messiah, \textit{Quantum Mechanics} (Dover, New York,
1999).

\bibitem{Landau} L. D. Landau, \textit{Phys. Z. Sowjetunion} \textbf{2}, 46
(1932).


\bibitem{Gell} M. Gell-Mann and F. Low, \textit{Phys. Rev.} \textbf{84}, 350
(1951).

\bibitem{Berry} M.V. Berry, \textit{Proc. R. Soc. London A} \textbf{392}, 45
(1984).

\bibitem{Simon} B. Simon, \textit{Phys. Rev. Lett.} \textbf{51}, 2167 (1983).

\bibitem{Farhi} Edward Farhi, Jeffrey Goldstone, Sam Gutmann, Joshua Lapan,
Andrew Lundgren, Daniel Preda , \textit{Science} \textbf{292}, 472 (2001).

\bibitem{Steffen} Matthias Steffen, Wim van Dam, Tad Hogg, Greg Breyta,
Isaac Chuang,\textit{Phys. Rev. Lett.} \textbf{90}, 067903 (2003).

\bibitem{Roland} Jeremie Roland and Nicolas J. Cerf,\textit{Phys. Rev. A}
\textbf{65}, 042308 (2002).

\bibitem{Marzlin1} Karl-Peter Marzlin, Barry C. Sanders, \textit{Phys. Rev.
Lett.} \textbf{93}, 160408 (2004).

\bibitem{Marzlin2} Solomon Duki, H. Mathur, Onuttom Narayan, \textit{Phys.
Rev. Lett.} \textbf{97}, 128901 (2006).

\bibitem{Marzlin3} Jie Ma, Yongping Zhang, Enge Wang, Biao Wu, \textit{Phys.
Rev. Lett.} \textbf{97}, 128902 (2006).

\bibitem{Marzlin4} Karl-Peter Marzlin, Barry C. Sanders, \textit{Phys. Rev.
Lett} \textbf{97}, 128903 (2006).

\bibitem{Wu1} Zhaoyan Wu and Hui Yang, \textit{Phys. Rev. A} \textbf{72},
012114 (2005).

\bibitem{Tong1} D. M. Tong, K. Singh, L. C. Kwek, C. H. Oh, \textit{Phys.
Rev. Lett.} \textbf{95}, 110407 (2005).

\bibitem{Tong2} D. M. Tong, K. Singh, L. C. Kwek, C. H. Oh, \textit{Phys.
Rev. Lett.} \textbf{98}, 150402 (2007); R. MacKenzie, E. Marcotte, H.
Paquette, \textit{Phys. Rev. A} \textbf{73}, 042104 (2006); Jian-Lan Chen,
Mei-sheng Zhao, Jian-da Wu and Yong-de Zhang, \textit{quant-ph/07060299};
Jian-da Wu, Mei-sheng Zhao, Jian-lan Chen, Yong-de Zhang, \textit{%
arXiv:0706.0264}.

\bibitem{Wei} Zhaohui Wei and Mingsheng Ying, \textit{Phys. Rev. A} \textbf{%
76}, 024304 (2007).

\bibitem{Sarandy} M. S. Sarandy and D. A. Lidar, \textit{Phys. Rev. Lett.}
\textbf{71}, 012331 (2005); M. S. Sarandy and D. A. Lidar, \textit{Phys.
Rev. Lett.} \textbf{95}, 250503 (2005); Han Pu, Peter Maenner, Weiping
Zhang, and Hong Y. Ling, \textit{Phys. Rev. Lett.} \textbf{98}, 050406
(2007).


\end{thebibliography}
\end{document}